\documentclass[twocolumn,showpacs,preprintnumbers,amsmath,amssymb,aps]{revtex4-2}
\usepackage{epsfig}
\usepackage{epstopdf}
\usepackage{graphicx}
\usepackage{dcolumn}
\usepackage{bm}
\usepackage{slashbox}
\usepackage{textcomp}
\usepackage{array}
\usepackage{subcaption}
\usepackage{booktabs}
\usepackage{gensymb}

\begin{document}

\title{Thermoelectric properties of marcasite-type compounds MSb$_2$ (M = Ta, Nb): A combined experimental and computational study}

\author{Shamim Sk$^{1}$}
\author{Naoki Sato$^{1,}$}
\altaffiliation{Electronic mail: SATO.Naoki@nims.go.jp}
\author{Takao Mori$^{1,}$}
\altaffiliation{Electronic mail: MORI.Takao@nims.go.jp}
\affiliation{$^{1}$Research Center for Materials Nanoarchitectonics (MANA), National Institute for Materials Science (NIMS), 1-1 Namiki, Tsukuba, Ibaraki 305-0044, Japan}


\begin{abstract}
Here, we investigate the thermoelectric properties of the marcasite-type compounds MSb$_2$ (M = Ta, Nb) in the temperature range of 310--730 K. These compounds were synthesized by a solid-state reaction followed by the spark plasma sintering process. The Rietveld refinement method confirms the monoclinic phase with space group \textit{C}2/\textit{m} for both compounds. The observed values of Seebeck coefficients exhibit non-monotonic behaviour in the studied temperature range, with the maximum magnitude of $-$14.4 and $-$22.7 $\micro$V K$^{-1}$ for TaSb$_2$ and NbSb$_2$, respectively at $\sim$444 K. The negative sign of \textit{S} in the full temperature window signifies the \textit{n}-type behaviour of these compounds. Both electrical and thermal conductivities show an increasing trend with temperature. The experimentally observed thermoelectric properties are understood through the first-principles DFT and Boltzmann transport equation. A pseudogap in the density of states around the Fermi level characterizes the semimetallic behaviour of these compounds. The multi-band electron and hole pockets were found to be mainly responsible for the temperature dependence of transport properties. The experimental power factors are found to be $\sim$0.09 and $\sim$0.42 mW m$^{-1}$ K$^{-2}$ at 310 K for TaSb$_2$ and NbSb$_2$, respectively. From the DFT-based calculations, the maximum possible power factors for \textit{p}-type conduction are predicted as $\sim$1.14 and $\sim$1.74 mW m$^{-1}$ K$^{-2}$, while these values are found to be $\sim$1.16 and $\sim$1.80 mW m$^{-1}$ K$^{-2}$ for \textit{n}-type TaSb$_2$ and NbSb$_2$, respectively at 300 K with the corresponding doping concentrations. The present study suggests that the combined DFT and Boltzmann transport theory are found to be reasonably good at explaining the experimental transport properties, and moderate power factors are predicted. 

\vspace{0.3cm}
Key words: Thermoelectric properties, density functional theory, electronic structure, electron and hole pockets, semi-
classical Boltzmann theory, power factor.

\end{abstract}

\maketitle
\section{INTRODUCTION}
The performance of thermoelectric (TE) materials is evaluated by the dimensionless figure-of-merit, $ZT=\frac{S^2\sigma T}{\kappa}$, where \textit{S} is the Seebeck coefficient, $\sigma$ is the electrical conductivity, $\kappa$ is the thermal conductivity (which is the sum of electronic, $\kappa_e$ and lattice thermal conductivity, $\kappa_l$), and \textit{T} is the absolute temperature. The value of \textit{ZT} determines the conversion efficiency; the higher the \textit{ZT} value, the better the TE performance. Finding materials having \textit{ZT} higher than unity is still a challenging task, though the research in TE has made in progress for many decades \cite{snyder2008,sootsman2009,zhao2014,mori2017,shittu2020,henricks2022,angelo2023}. Actually, efficient TE materials have to pass through a typical tradeoff, which includes materials that are good electrical conductors but have poor thermal conductivity. This implies that the transport of charge carriers within the material should experience weak electron scattering and strong phonon scattering.

Realizing high \textit{ZT} has always been a challenging task due to the strong correlation among the TE parameters through charge carriers \cite{ashcroft,shamim_mrx}. Till now, the state-of-the-art TE materials are Bi$_2$Te$_3$ \cite{bite1,bite2}, Sb$_2$Te$_3$ \cite{sbte}; PbTe \cite{pbte1,pbte2}, SnTe \cite{snte}; and SiGe \cite{sige1,sige2} based compounds. Recently, the antimonides have gained increased interest with high performance discovered in Mg$_3$Sb$_2$-type materials for example \cite{tamaki2016,liu2021,liu2022,wand2023}. The expression of \textit{ZT} implies that there are two ways for boosting the TE performance: either by decreasing the lattice thermal conductivity without affecting or less affecting the electronic properties \cite{snyder2008,wang2013,shi2018,khan} or by improving the power factor ($S^2\sigma$) \cite{mahan1996,ahmed2017,liu2020,bourges2020,garmroudi2022,zhang2023}.

In recent decades, the TE properties of FeSb$_2$ were extensively studied because of ultra-high \textit{S} of up to $-$45000 $\micro$V K$^{-1}$ at 12 K, resulting in the highest power factor reported ever \cite{fesb1,fesb2}. But, due to the large $\kappa$ of FeSb$_2$, the \textit{ZT} is diminished with the low value of 0.005 at 12 K \cite{fesb1,fesb2}. The FeSb$_2$ is a narrow-gap semiconductor with an orthorhombic marcasite structure \cite{fesb1,fesb2}. The TE properties of other marcasite-type compounds including XTe$_2$ (X = Fe, Co, Ni) \cite{bai}, FeX$_2$ (X = Se, Te) \cite{gudelli2014}, FeS$_2$ \cite{gudelli2013}, FeAs$_2$ \cite{sun1}, RuSb$_2$ \cite{sun1,sun2}, etc, have been studied in the last decades. In the same group of marcasite-type compounds, the MSb$_2$ (M = Ta, Nb) with monoclinic crystal structure (Space group: \textit{C}2/\textit{m}, No. 12) have recently attracted remarkable attention as Weyl semimetals having unexpected magneto-transport properties \cite{tasb1,tasb2,tasb3,nbsb1,nbsb2}. Regrettably, the TE properties of these compounds are rarely explored, especially at elevated temperatures. For instance, Masuda \textit{et al.} have studied the TE properties of TaSb$_2$ compound in the temperature range of 300--800 K \cite{masuda}. They synthesized TaSb$_2$ using solid-state reaction method with the spark plasma-sintering process and explored the temperature dependent \textit{S} and $\sigma$ \cite{masuda}. The experimental measurements of \textit{S} and $\sigma$ of MSb$_2$ (M = Ta, Nb) have been done by Failamani \textit{et al.} \cite{failamani} in the temperature range of 300--800 K. However, there is a lack of theoretical understanding of the experimental TE properties of MSb$_2$ (M = Ta, Nb) at elevated temperatures. In the present study, the TE properties of MSb$_2$ (M = Ta, Nb) are investigated using both experimental and computational tools at the high temperatures. The experimental TE properties of said compounds are understood using density functional theory (DFT)-based calculations, and we predict the possible maximum power factors with a suitable amount of doping.  

In this work, we have synthesized marcasite-type compounds MSb$_2$ (M = Ta, Nb) using the combined solid-state reaction and spark plasma sintering process. All the TE properties are measured and analyzed. First-principle calculations and Boltzmann transport equations are utilized to understand the experimental results. The multi-band electron and hole pockets are found to reasonably explain the experimental data. The maximum possible power factors for \textit{p}-type and \textit{n}-type of MSb$_2$ (M = Ta, Nb) are also predicted using the DFT-based calculations. 

\section{EXPERIMENTAL AND COMPUTATIONAL DETAILS}
MSb$_2$ (M = Ta, Nb) were synthesized using a combined solid state reaction (SSR) and spark plasma sintering (SPS) process. The high purity powder of Ta (99.99$\%$), Nb (99.99$\%$) and Sb (99.99$\%$) from Sigma-Aldrich were taken as starting materials. The desired amount of metal powder were ground in alumina mortar and cold pressed at 25 MPa to form pellets. Then the obtained pellets were heated at 700 $\celsius$ for 3 days in vacuum-sealed quartz tube. Although the target phase for both compounds were obtained in SSR, but the pellets were too brittle for the transport measurements. In order to get dense pellets, we ground the SSR pellets and did the SPS (SPS-1080, SPS Syntex Inc.) at 750 $\celsius$ for 8 minutes. During the SPS, the constant uniaxial pressure of 50 MPa was applied to a graphite punch of 10 mm diameter under a partial argon atmosphere. The relative densities were calculated from the sample densities determined by Archimedes principle, which are found to be $\sim$95$\%$ and $\sim$94$\%$ for TaSb$_2$ and NbSb$_2$, respectively. The obtained pellets were cut into the required dimensions for the transport measurements.

The simultaneous measurement of the electrical conductivity and Seebeck coefficient were performed using ZEM-2 (ADVANCE RIKO) under partial helium environment. The dimensions of TaSb$_2$ and NbSb$_2$ were taken as $4.11\times1.92\times7.57$ mm$^3$ and $4.11\times1.62\times7.39$ mm$^3$, respectively. 

The thermal conductivity was obtained using the formula: $\kappa=D\times C_p\times\rho$. The $\rho$ is the density of the sample. The \textit{D} is the thermal diffusivity, which is measured using the laser flash diffusivity method implemented in LFA-467 Hyper flash (Netzsch) instrument. The heat capacity, $C_p$ was estimated using a standard sample (pyroceram-9060) in LFA-467. The circular pellets of TaSb$_2$ and NbSb$_2$ with diameter of 10 mm and thickness of 1.3 mm and 1.6 mm, respectively were used for the measurements.  

\begin{figure}
\includegraphics[width=0.9\linewidth, height=9.5cm]{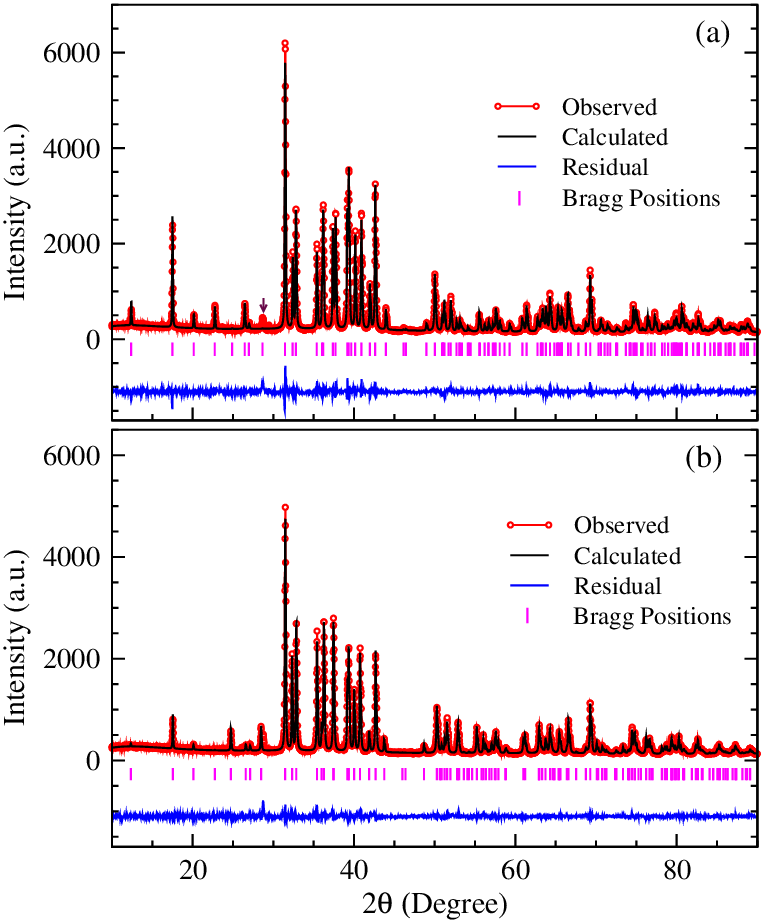} 
\caption{\small{Measured room temperature X-ray diffraction patterns and calculated ones by Rietveld method of (a) TaSb$_2$ and (b) NbSb$_2$. The peak at 28.7$\degree$ indicated by an arrow in (a) is an impurity peak.}}
\end{figure}  

\begin{table*}
\caption{\label{tab:table1}%
\small{Fractional coordinates of atomic positions for relaxed structure of TaSb$_2$ and NbSb$_2$. The experimental values are also mentioned in the parentheses.}}

\begin{ruledtabular}
\begin{tabular}{lccccccc}
\textrm{TaSb$_2$}&
\textrm{x (exp)}&
\textrm{y (exp)}&
\textrm{z (exp)}&
\textrm{NbSb$_2$}&
\textrm{x (exp)}&
\textrm{y (exp)}&
\textrm{z (exp)}\\
      
\colrule
Ta        & 0.150 (0.152) & 0 (0) & 0.188 (0.193) & Nb & 0.151 (0.152) & 0 (0) & 0.190 (0.194) \\
Sb1       & 0.148 (0.143) & 0 (0) & 0.535 (0.532) & Sb1 & 0.149 (0.145) & 0 (0) & 0.536 (0.531) \\
Sb2       & 0.405 (0.401) & 0 (0) & 0.114 (0.111) & Sb2 & 0.405 (0.402) & 0 (0) & 0.115 (0.111) \\
\end{tabular}
\end{ruledtabular}
\end{table*}

X-ray diffraction (XRD) were taken (Smart Lab3, Rigaku) in 10--90$\degree$ as shown in Fig. 1(a) and (b) for TaSb$_2$ and NbSb$_2$, respectively. Rietveld refinement method confirms the monoclinic phase with space group \textit{C}2/\textit{m} (No. 12) for both compounds. The refined lattice parameters are obtained as $a=10.22$ \AA \,, $b=3.64$ \AA \,, $c=8.29$ \AA \, and $\beta=120.39\degree$ for TaSb$_2$, and $a=10.23$ \AA \,, $b=3.63$ \AA \,, $c=8.33$ \AA \, and $\beta=120.02\degree$ for NbSb$_2$. The impurity peak at 28.7$\degree$ (Fig. 1(a)) may arise from the surface oxide layer of Sb$_2$O$_3$ \cite{masuda}.

In order to understand the experimental transport properties, we have carried out the ground state electronic structure calculations within (DFT) \cite{dft}. The projector augmented-wave method is used as implemented in Quantum Espresso code \cite{qe}. The PERDEW-ZUNGER (LDA) \cite{lda} exchange-correlation (XC) functional is used for the calculation. The calculations are done in relaxed structure with optimized lattice parameters of $a=10.14$ \AA \,, $b=3.62$ \AA \,, $c=8.22$ \AA \, and $\beta=120.53\degree$ for TaSb$_2$, and $a=10.16$ \AA \,, $b=3.59$ \AA \,, $c=8.25$ \AA \, and $\beta=120.01\degree$ for NbSb$_2$. The force convergence criteria for structure relaxation was set to be 10$^{-4}$ Ry/Bohr. Table 1 shows the relaxed atomic positions along with the experimental values for both compounds. The structures relaxed using DFT are closely matched with the experimental ones. The kinetic energy cut-off for wavefunctions is used as 60 Ry. The kinetic energy cut-off for charge density is set to be 8 times the kinetic energy cut-off for wavefunctions. The \textit{k}-mesh grid was used as $7\times15\times7$ for both compounds. The energy convergence criteria was set to be 10$^{-8}$ Ry for the self-consistent field calculation. The transport coefficients were calculated using the BoltzTraP2 package \cite{boltztrap2} interfaced with Quantum Espresso code \cite{qe}.

\section{RESULTS AND DISCUSSION}        
\subsection{EXPERIMENTAL TRANSPORT PROPERTIES}
Fig. 2(a) exhibits the experimentally measured Seebeck coefficients (\textit{S}) of MSb$_2$ (M = Ta, Nb) in the temperature range of 310--730 K. The $|S|$ for both compounds are found to increase up to $\sim$450 K, then decreases till the highest temperature, consistent with the other reported data \cite{masuda,failamani}. The highest magnitudes of \textit{S} are found to be $-$14.4 and $-$22.7 $\micro$V K$^{-1}$ for TaSb$_2$ and NbSb$_2$, respectively at $\sim$444 K. The total \textit{S} mainly comes from the contributions of electrons and holes. Under the two-carrier conduction model, the \textit{S} can be expressed as  \cite{ioffe} $S = \frac{S_{h}\sigma_{h}+S_{e}\sigma_{e}}{\sigma_{h}+\sigma_{e}}$, where $S_{h,e}$ is the Seebeck coefficients of holes and electrons and $\sigma_{h,e}$ is the electrical conductivity of holes and electrons. The sign of the \textit{S} is determined by the major contribution coming from the electrons or holes, because electrons generally yield the negative \textit{S}, while holes give positive \textit{S}. Therefore, the negative sign of \textit{S} in Fig. 2(a) signifies the dominating \textit{n}-type behaviour of these compounds. The magnitude of \textit{S} for NbSb$_2$ is larger than that of TaSb$_2$ in the full temperature window. The non-linear behaviour of \textit{S} with temperature for both compounds can be explained by the calculated band-structure, which is discussed later. 

\begin{figure*}
\begin{subfigure}{0.32\textwidth}
\includegraphics[width=0.9\linewidth, height=4.2cm]{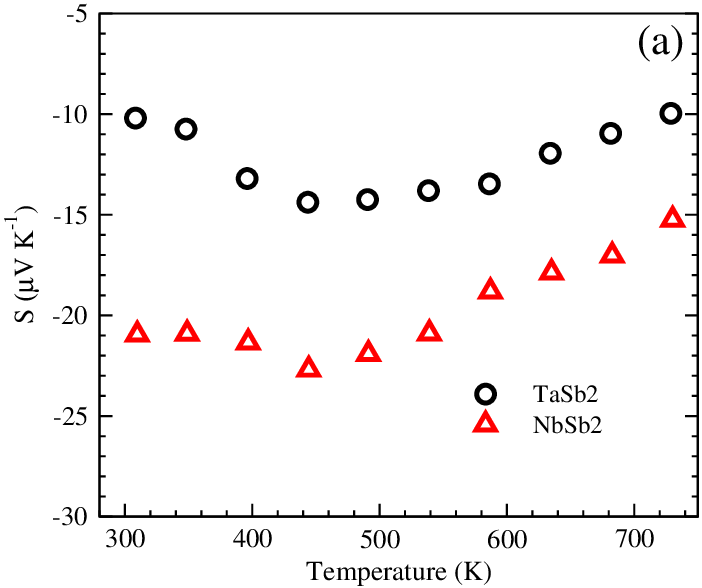} 
\end{subfigure}
\begin{subfigure}{0.32\textwidth}
\includegraphics[width=0.9\linewidth, height=4.2cm]{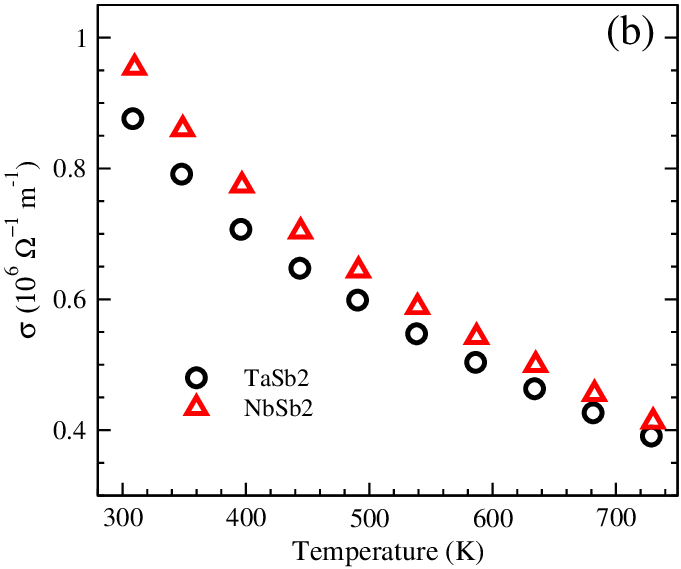} 
\end{subfigure}
\begin{subfigure}{0.32\textwidth}
\includegraphics[width=0.88\linewidth, height=4.2cm]{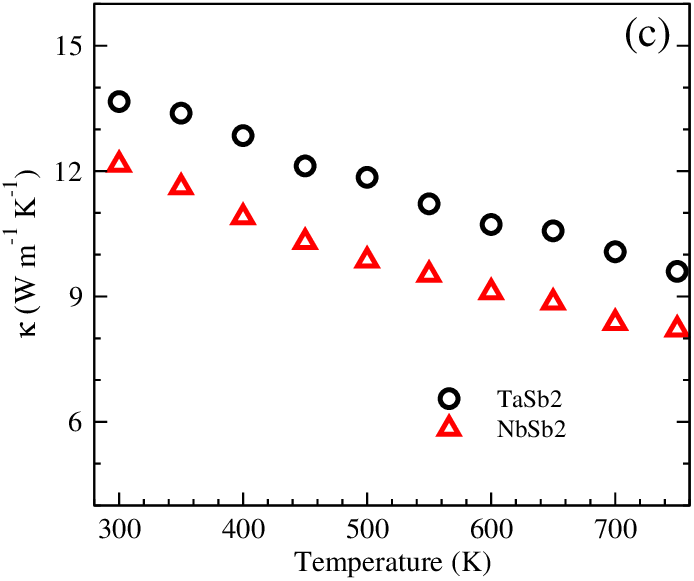} 
\end{subfigure}
\caption{\small{Temperature dependence of measured (a) Seebeck coefficient, \textit{S} (b) electrical conductivity, $\sigma$ and (c) thermal conductivity, $\kappa$ of TaSb$_2$ and NbSb$_2$.}}
\end{figure*}

\begin{figure*}
\begin{subfigure}{0.48\textwidth}
\includegraphics[width=0.97\linewidth, height=4.2cm]{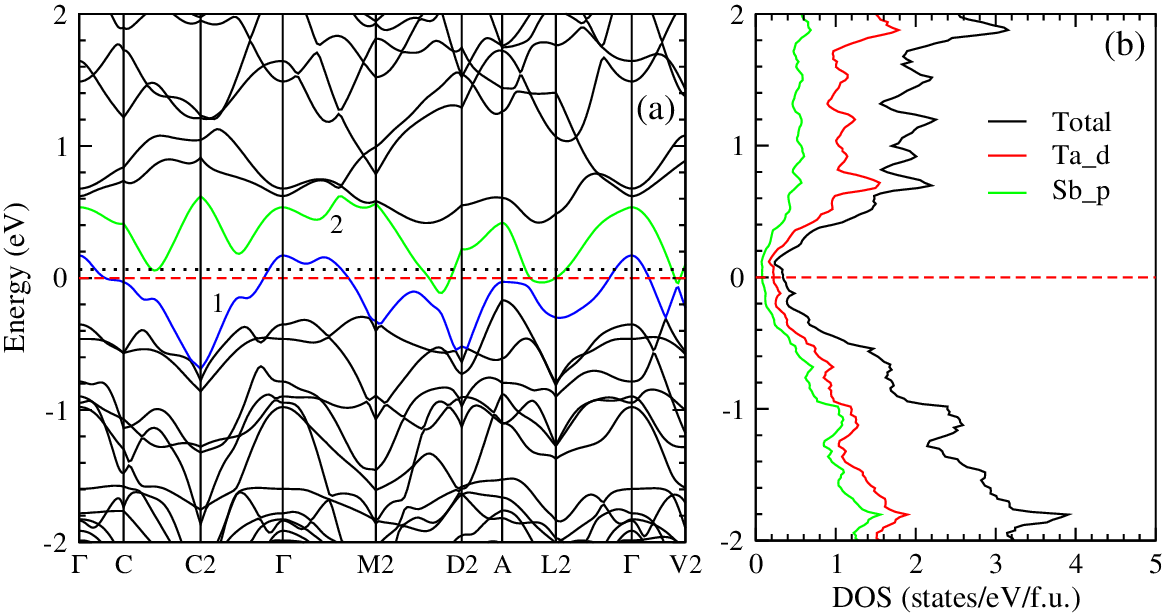} 
\end{subfigure}
\begin{subfigure}{0.48\textwidth}
\includegraphics[width=0.97\linewidth, height=4.2cm]{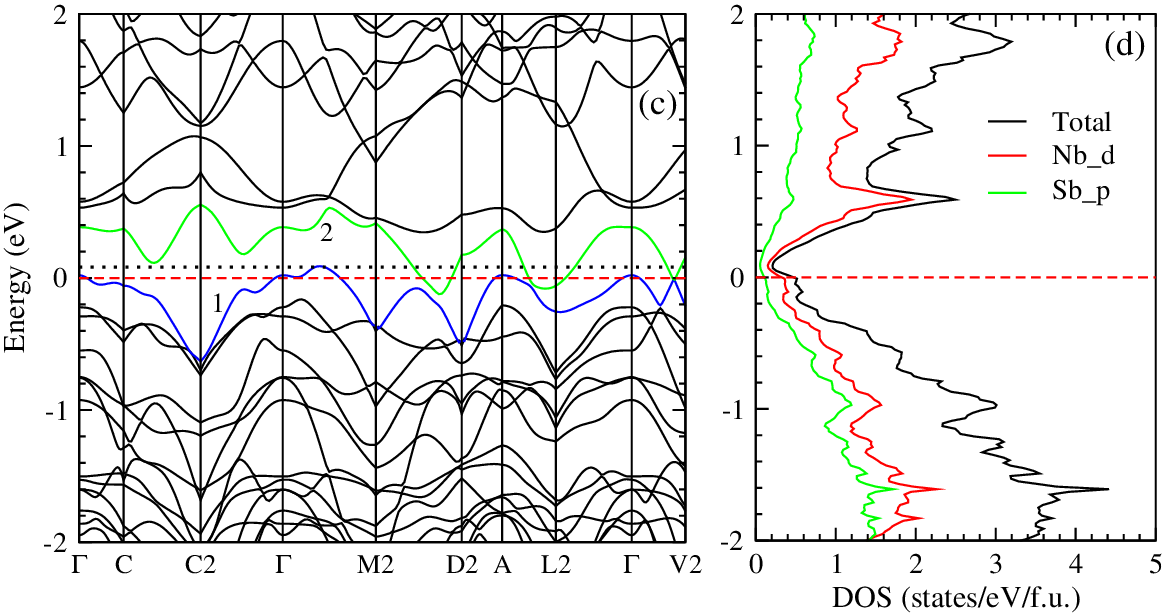} 
\end{subfigure}
\caption{\small{(a) Band-structure and (b) density of states (DOS) of TaSb$_2$; (c) band-structure and (d) DOS of NbSb$_2$.}}
\end{figure*}

Fig. 2(b) shows the temperature dependence of electrical conductivities ($\sigma$) of TaSb$_2$ and NbSb$_2$. The $\sigma$ are found to be 0.88 $\times 10^{6}$ and 0.95 $\times 10^{6}$ $\ohm^{-1}$ m$^{-1}$ for TaSb$_2$ and NbSb$_2$, respectively at $\sim$310 K. Then as the temperature increases, the $\sigma$ decreases till the highest temperature ($\sim$730 K) with the corresponding values of 0.39 $\times 10^{6}$ and 0.41 $\times 10^{6}$ $\ohm^{-1}$ m$^{-1}$ for TaSb$_2$ and NbSb$_2$, respectively. Our measured values of $\sigma$ are accordance with the other experimentally reported data \cite{masuda,failamani}. The decrement behaviour of $\sigma$ can be explained by the simple formula: $\sigma = \frac{ne^{2}\tau}{m^{*}}$, where, $n$, $e$, $\tau$ and $m^{*}$ are the carrier concentration, the electronic charge, the relaxation time and the effective mass of charge carriers, respectively. Under the increment of temperature, $n$ always increases, whereas $\tau$ decreases for any compound. Among the opposite trend of $n$ and $\tau$ with temperature, the dominating behaviour generally gives the temperature dependent trend of $\sigma$. Hence, Fig. 2(b) shows that the temperature dependent $\tau$ is dominating over $n$ in $\sigma$. In general, with increase in temperature, the $\sigma$ of semiconductors increases, while $\sigma$ of metals decreases. Hence, the temperature dependent trend of $\sigma$ of present compounds show the metal-like behaviour. However, the electronic structure calculations predict the semimetallic (in between semiconductor and metals) behaviour of these compounds, which is discussed later.

Using the measured \textit{S} and $\sigma$, the power factors are calculated as $\sim$0.91 $\times$ 10$^{-1}$ and $\sim$4.17 $\times$ 10$^{-1}$ mW m$^{-1}$ K$^{-2}$ at 310 K for TaSb$_2$ and NbSb$_2$, respectively. At 730 K, the power factors are calculated as $\sim$0.39 $\times$ 10$^{-1}$ and $\sim$2.32 $\times$ 10$^{-1}$ mW m$^{-1}$ K$^{-2}$, respectively. Usually, semimetals possess low \textit{S} with high $\sigma$. But recent studies have shown that the \textit{S} of the semimetals can be tuned by magnetic field, strain, etc \cite{kaur,skinner,pan_2021}. Semimetals having asymmetry in their electron and hole pockets could have high \textit{S} \cite{markov,mao}. 

The thermal conductivities ($\kappa$) of TaSb$_2$ and NbSb$_2$ are measured in the temperature region 300--750 K as shown in Fig. 2(c). The $\kappa$ are gradually decreasing with the increment of temperature in the whole temperature range. At 300 K, the $\kappa$  are found to be 13.7 and 12.2 W m$^{-1}$ K$^{-1}$, while these are decreased to 9.6 and 8.2 W m$^{-1}$ K$^{-1}$ at 750 K for TaSb$_2$ and NbSb$_2$, respectively. In the full temperature window, the values of $\kappa$ for TaSb$_2$ are higher than the NbSb$_2$. The total $\kappa$ is a simple addition of electronic thermal conductivity ($\kappa_{e}$) and lattice thermal conductivity ($\kappa_{L}$), \textit{i.e.}, $\kappa=\kappa_{e}+\kappa_{L}$. In this work, we have calculated the temperature dependence of $\kappa_{e}$ for both compounds. The experimental $\kappa_{e}$ can be estimated using the experimental $\sigma$ via Wiedemann-Franz law: $\kappa_{e}=L\sigma T$, \textit{L} is Lorenz number. Then it will be interesting to see how calculated $\kappa_{e}$ explains the experimental $\kappa_{e}$, which is described later. As can be seen from the Fig. 2(b), the $\sigma$ of NbSb$_2$ are higher than that of TaSb$_2$. Hence, larger values of $\kappa_{e}$ are expected for NbSb$_2$ as compared to TaSb$_2$ according to the Wiedemann-Franz law.

\subsection{ELECTRONIC STRUCTURE}
To understand the experimentally measured transport coefficients, we have calculated the electronic structure of MSb$_2$ (M = Ta, Nb). The band-structures for TaSb$_2$ and NbSb$_2$ are calculated along the lines between high symmetry points ($\Gamma$--C--C2--$\Gamma$--M2--D2--A--L2--$\Gamma$--V2) in the first Brillouin zone, which are shown in Fig. 3(a) and (c), respectively. The dashed red line corresponding to zero energy defines the Fermi level, $E_{F}$ of the compounds. It is clear that the two bands (which are indexed by 1 and 2) around the $E_{F}$ are expected to contribute to the transport properties of these compounds. In both Fig. 3(a) and (c), the occupied band 1 crosses the $E_{F}$ at either side of the $\Gamma$ point and becomes unoccupied, while the unoccupied band 2 crosses the $E_{F}$ in the M2--D2, A--L2 and $\Gamma$--V2 directions and becomes occupied. This type of mixing of occupied and unoccupied bands around the $E_{F}$ predicts the semimetal-like character of the compound, which is consistent with the other reported works \cite{tasb1,tasb2,nbsb1,lee2021}. The electronic band-structure is the key input for calculating any electronic transport properties. In the present study, we have calculated the Seebeck coefficient, electrical conductivity and electronic part of thermal conductivity for MSb$_2$ (M = Ta, Nb). We will recall this part in the next sub-section during the discussion of calculated transport properties. 

Fig. 3(b) and (d) express the calculated total and partial density of states (DOS) of TaSb$_2$ and NbSb$_2$, respectively. At $E_{F}$, the values of DOS are calculated as $\sim$0.34 and $\sim$0.47 states/eV/f.u. for TaSb$_2$ and NbSb$_2$, respectively. A pseudogap around the $E_{F}$ characterizes the semimetallic behaviour of these compounds. In order to know the contributions in transport properties from different atoms, we have calculated the partial DOS of Ta, Nb and Sb as shown in the same figures. In the occupied band region, the contribution in the DOS comes from Ta(Nb)-\textit{d} and Sb-\textit{p} orbitals almost equally. In contrast, in the unoccupied band region, the dominant contribution in the DOS comes from Ta(Nb)-\textit{d} orbitals, with the small contribution from Sb-\textit{p} orbitals. In the energy range of $-1$ to 0 eV in the occupied band region of Fig. 3(b), the contributions of Ta-\textit{d} and Sb-\textit{p} in the DOS are calculated as $\sim$60$\%$ and $\sim$40$\%$, while these contributions are found to be $\sim$80$\%$ and $\sim$20$\%$, respectively, in the energy range of 0 to 1 eV in the unoccupied band region.

\begin{figure}
\includegraphics[width=0.8\linewidth, height=5.3cm]{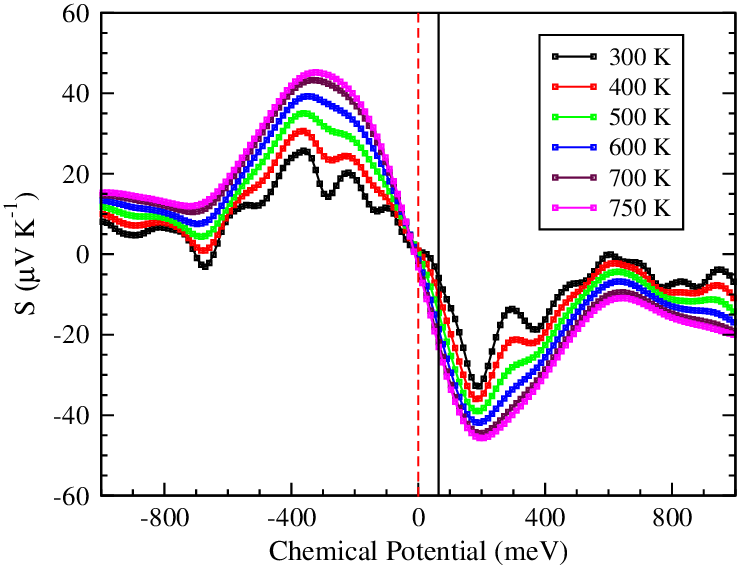} 
\caption{\small{Calculated Seebeck coefficient, \textit{S} at the function of chemical potential, $\mu$ at different temperatures of TaSb$_2$. The solid vertical black line indicates the $\mu$ value, where transport properties are calculated.}}
\end{figure}

\subsection{CALCULATED TRANSPORT PROPERTIES}
In this sub-section, we discuss the calculated transport properties, \textit{viz.,} the temperature dependence of Seebeck coefficient (\textit{S}), electrical conductivity divided by relaxation time ($\sigma/\tau$) and electronic thermal conductivity divided by relaxation time ($\kappa_{e}/\tau$) as calculated using the BoltzTraP2 package \cite{boltztrap2} under semi-classical Boltzmann theory. The BoltzTrap2 is based on the combined constant relaxation time approximation (CRTA) and rigid band approximation (RBA). In CRTA, relaxation time ($\tau$) is considered as a constant, but in principle $\tau$ is dependent on both band index and wave vector direction. The RBA means that the band-structure is independent of temperature and doping.  

At first, we have calculated the \textit{S} at E$_{F}$, which are $\sim$1 and $\sim$20 $\micro$V K$^{-1}$ at 310 K for TaSb$_2$ and NbSb$_2$, respectively. These positive values of \textit{S} are far away from the experimental negative values of $\sim-10$ and $\sim-21$ $\micro$V K$^{-1}$ for TaSb$_2$ and NbSb$_2$, respectively at the same temperature. At this point, it is important to note that the calculations of \textit{S} have been done on single crystalline stoichiometric compounds. But, it is very common to have off-stoichiometry in any synthesized polycrystalline samples. This off-stoichiometry may come from many factors, including the purity of the starting materials, inaccuracy in weighing the raw materials, inhomogeneous mixing during the synthesis process, evaporation of low melting element during the heat treatment etc. These factors are mainly responsible for the defects and/or disorders in the samples which may affect the \textit{S} of the sample. In addition to this, the anisotropy often matters when comparing the calculation of single crystal with the experimental polycrystalline one. All these factors can be addressed in the calculation by shifting the chemical potential ($\mu$) of the compound. But, quantifying the exact value of $\mu$ is a challenging job for any compound. For doing this, we have calculated $\mu$ dependent \textit{S} at different temperatures as shown in Fig. 4. Then, the $\mu$ has been chosen at 300 K for a better representation of experimental \textit{S}. We found that at $\mu$ $\approx$ 64 and 82 meV, the calculated \textit{T} dependent \textit{S} gives the best match with the experimental \textit{T} dependent \textit{S} for TaSb$_2$ and NbSb$_2$ as shown in Fig. 5 (a) and (b), respectively. This constant $\mu$ calculated at 300 K is used to calculate the other transport properties in the full temperature range.

\begin{figure}
\includegraphics[width=0.78\linewidth, height=8.8cm]{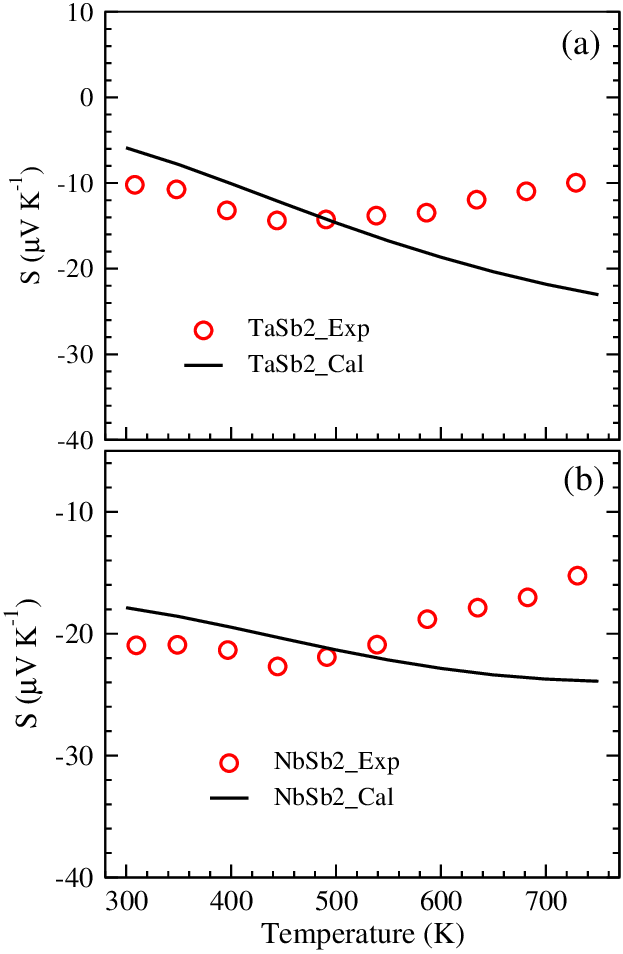} 
\caption{\small{Comparison of experimental and calculated Seebeck coefficients, \textit{S} of (a) TaSb$_2$ and (b) NbSb$_2$.}}
\end{figure}

The \textit{T} dependent \textit{S} of TaSb$_2$ and NbSb$_2$ at $\mu$ $\approx$ 64 and 82 meV, respectively can be understood through electron/hole pockets formed with the bands 1 and 2 in the band-structure of Fig. 3(a) and (c). The black horizontal dotted lines above the $E_F$ indicate the $\mu$ values, where the temperature dependent \textit{S} are calculated. For both compounds, it is clear that in \textit{S}, the contributions of charge carriers mainly come from the hole pockets formed with band 1 at the vicinity of $\Gamma$ and A points and in the $\Gamma$--M2 direction, and electron pockets formed with the band 2 in the C--C2, C2--$\Gamma$ M2--D2, A--L2, $\Gamma$--V2 and $\Gamma$--M2 (band 1) directions at $\mu$ $\approx$ 64 (82) meV. The presence of dominating electron pockets over hole pockets gives the negative \textit{S} for both compounds. The size of the hole pockets at $\Gamma$ point for TaSb$_2$ (at 64 meV) is larger than the hole pocket at $\Gamma$ point for NbSb$_2$ (at 82 meV), which supports the less magnitude of the \textit{S} for TaSb$_2$ as compared to that of NbSb$_2$. However, in the high temperature region, the calculated \textit{S} are deviating from the experimental \textit{S}. At this conjuncture, it is important to note that the \textit{S} is calculated using the ground-state band-structures and constant $\mu$ of 300 K. But, the band-structure and $\mu$ are the temperature dependent quantities. Hence, the temperature dependence of these quantities are expected to improve the accuracy of calculated \textit{S} at higher temperature.

\begin{figure}
\includegraphics[width=0.75\linewidth, height=9.0cm]{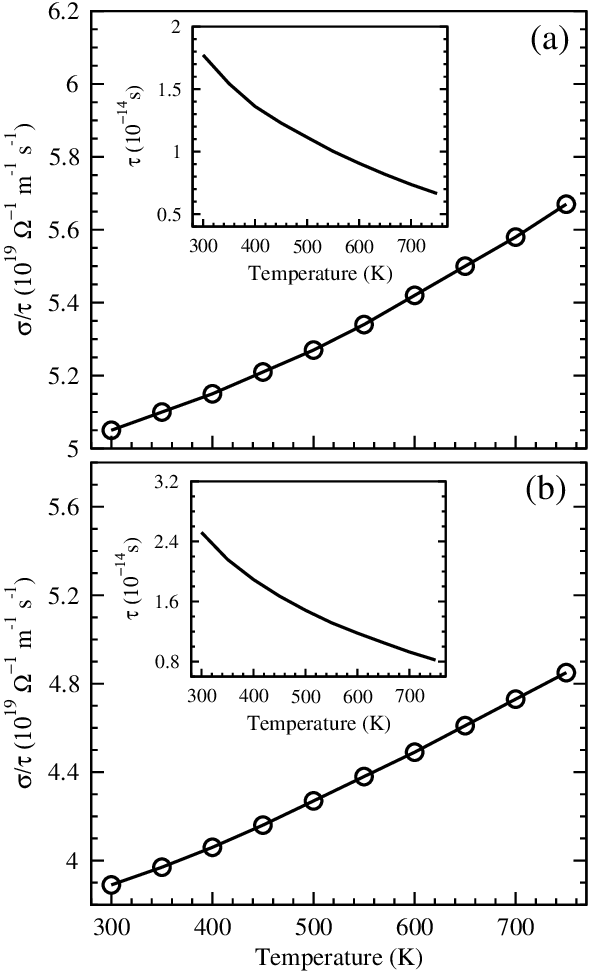} 
\caption{\small{Calculated electrical conductivity divided by relaxation time, $\sigma/\tau$ of (a) TaSb$_2$ and (b) NbSb$_2$. Inset shows $\tau$ as a function of temperature, which is estimated by comparing the calculated $\sigma/\tau$ with experimental $\sigma$.}}
\end{figure} 

Fig. 6 shows the calculated $\boldsymbol{\sigma}/\tau$ of TaSb$_2$ and NbSb$_2$ in the temperature range of 300--750 K. The $\boldsymbol{\sigma}$ of $n^{th}$ band is defined as \cite{ashcroft}: 
\begin{equation}
\boldsymbol{\sigma}^{(n)} = e^{2}\int\frac{d\textbf{k}}{4\pi^{3}}\tau_{n}(\varepsilon_{n}(\textbf{k}))\textbf{v}_{n}(\textbf{k})\textbf{v}_{n}(\textbf{k})\bigg(-\frac{\partial f}{\partial \varepsilon}\bigg)_{\varepsilon=\varepsilon_{n}(\textbf{k})},
\end{equation}
where $e$ is an electronic charge, $\tau_{n}(\varepsilon_{n}(\textbf{k}))$ is the relaxation time of an electron of $n^{th}$ band with wave vector \textbf{k} and $\varepsilon_{n}(\textbf{k})$ is an energy band. The $\textbf{v}_{n}(\textbf{k})$ is the mean velocity of an electron of $n^{th}$ band with wave vector \textbf{k}. The \textit{f} is the Fermi-Dirac distribution function, which takes care of the temperature dependency of the compound. As we discussed earlier, the BoltzTraP2 code \cite{boltztrap2} works under constant relaxation time approximation, \textit{i.e.,} $\tau_{n}(\varepsilon_{n}(\textbf{k})) = \tau$. Hence, the calculated $\boldsymbol{\sigma}/\tau$ depends on $\textbf{v}_{n}(\textbf{k})$ and $\frac{\partial f}{\partial \varepsilon}$, and number of available states at a given $\mu$. With an increase in temperature, the number of states always increases. Hence, the increasing nature of $\boldsymbol{\sigma}/\tau$ with temperature (Fig. 6) is directly related to the more number of available states at high temperature. Initially, the charge carriers from band 1 at the vicinity of $\Gamma$ point and in the $\Gamma$--M2 direction, and band 2 in the C--C2, M2--D2, A--L2 and $\Gamma$--V2 directions are participated in the transport (Fig 3(a)). As the temperature increases, the more charge carriers from band 1 at the vicinity of A point and in the M2--D2 and $\Gamma$--V2 directions; the band 2 in C2--$\Gamma$ direction are expected to contribute in the $\boldsymbol{\sigma}/\tau$. At high temperature, more charge carriers from the different electron/hole pockets participate in the conduction, and hence $\boldsymbol{\sigma}/\tau$ increases. The calculated temperature dependent $\boldsymbol{\sigma}/\tau$ is compared with the experimental $\sigma$ (Fig. 2(b)) to extract the temperature dependent $\tau$. Insets of Fig. 6 shows the extracted $\tau$ in the temperature range of 300--750 K. The values of $\tau$ are calculated as $\sim$1.8 $\times$ 10$^{-14}$ s and $\sim$2.5 $\times$ 10$^{-14}$ s at 300 K for TaSb$_2$ and NbSb$_2$, respectively. With the increase in temperature, $\tau$ decreases monotonically and reaches $\sim$0.66 $\times$ 10$^{-14}$ s ($\sim$0.82 $\times$ 10$^{-14}$ s) at 750 K. The decreasing nature of $\tau$ is due to the presence of more scattering centers at high temperature. For a real system, the calculation of $\tau$ is a challenging task due to the involvement of many scattering mechanisms, including electron-electron scattering, electron-phonon scattering, electron-defect scattering etc \cite{ashcroft}. This is the reason why we have chosen the simple method to extract temperature dependent $\tau$. It will be interesting to see how this temperature dependent $\tau$ explains the other transport properties.

\begin{figure}
\includegraphics[width=0.75\linewidth, height=9.0cm]{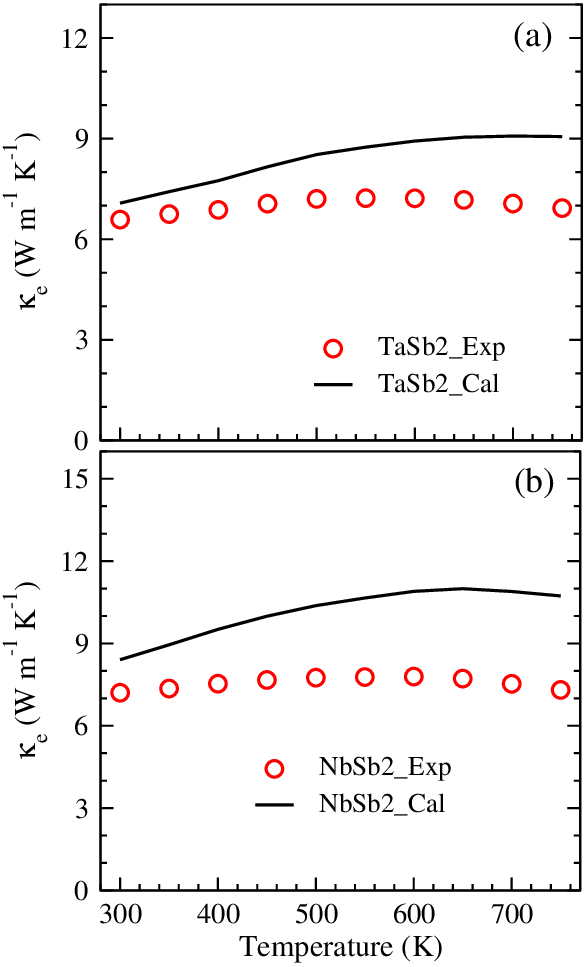} 
\caption{\small{Comparison of experimental (estimated using Wiedemann-Franz law) and calculated electronic part of thermal conductivity, $\kappa_{e}$ of (a) TaSb$_2$ and (b) NbSb$_2$.}}
\end{figure} 

\begin{figure*}
\begin{subfigure}{0.48\textwidth}
\includegraphics[width=0.95\linewidth, height=4.8cm]{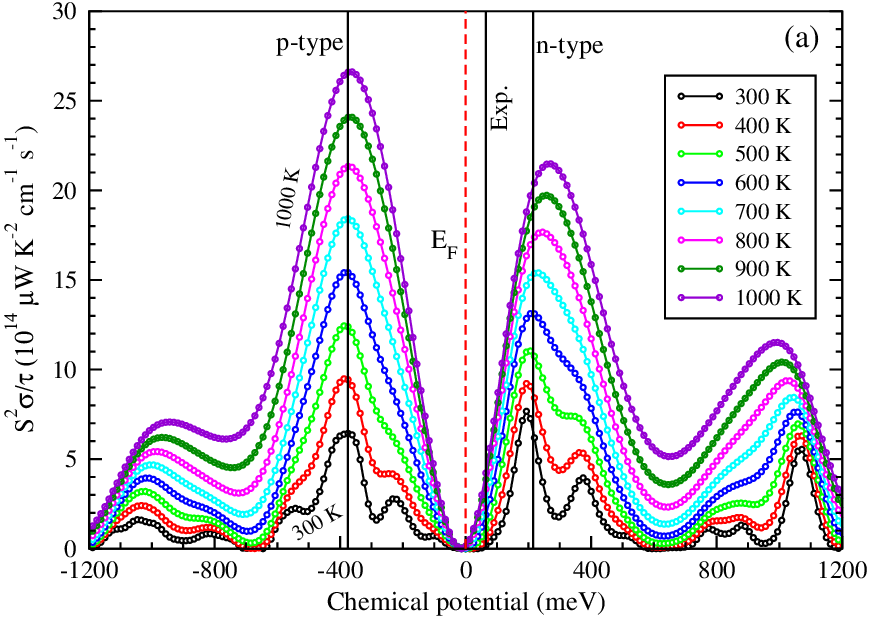} 
\end{subfigure}
\begin{subfigure}{0.48\textwidth}
\includegraphics[width=0.95\linewidth, height=4.8cm]{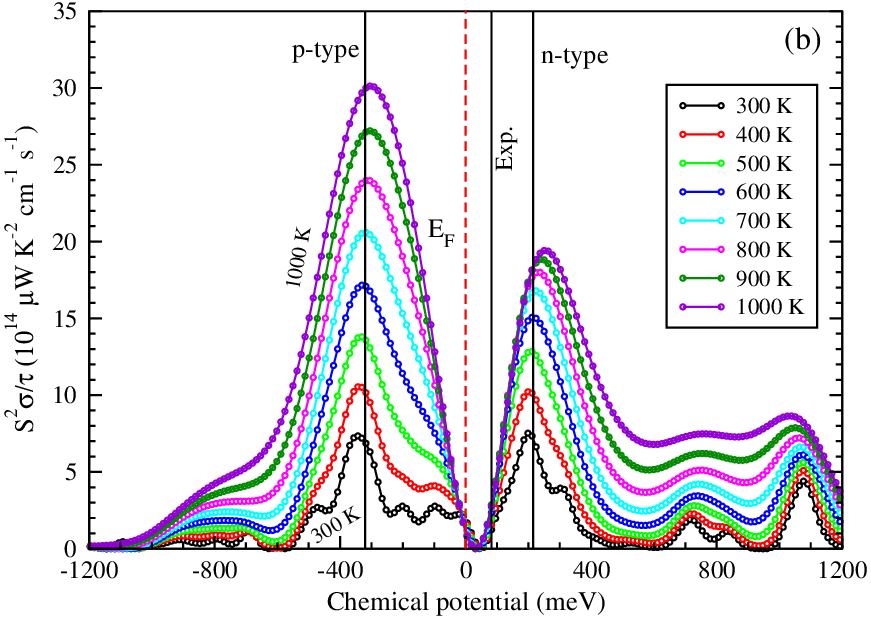} 
\end{subfigure}
\caption{\small{Variation of power factor ($S^2\sigma$) with chemical potential at different temperatures of (a) TaSb$_2$ and (b) NbSb$_2$.}}
\end{figure*}

We have calculated $\kappa_{e}/\tau$ for MSb$_2$ (M = Ta, Nb) using the following equation \cite{ashcroft}
\begin{equation}
\boldsymbol{\kappa_{e}}/\tau=\frac{\pi^{2}}{3}\bigg(\frac{k_{B}}{e}\bigg)^{2}T(\boldsymbol{\sigma/\tau}),
\end{equation}            
where $k_{B}$ in the Boltzmann constant. Then the $\kappa_{e}$ is computed using the temperature dependent $\tau$. Fig. 7 shows the $\kappa_{e}$ for TaSb$_2$ and NbSb$_2$ in the temperature region 300--750 K. The calculated $\kappa_{e}$ is compared with the experimental $\kappa_{e}$ in the same figure. The experimental $\kappa_{e}$ is estimated using Wiedemann-Franz law: $\kappa_e = L\sigma T$. Where the temperature dependent experimental $\sigma$ (Fig. 2(b)) and the constant value of \textit{L} (2.45 $\times$ 10$^{-8}$ W $\ohm$ K$^{-2}$) are taken to estimate the experimental $\kappa_{e}$. It is observed that the calculated $\kappa_{e}$ gives quite good agreement with the experimental $\kappa_{e}$. With an increase in temperature, $\kappa_{e}$ increases due to contributions of more charge carriers from the different electron/hole pockets. The figure shows that there is a small deviation at high temperature region. At this point, it is important to note that the calculations have been done using ground-state band-structure, constant $\mu$ and \textit{L} values, as we mentioned earlier. Considering temperature dependency of all these factors may improve the reproducibility at high temperature, which demands extra computational costs. 

\subsection{POWER FACTOR}
Finally, the chemical potential dependence of the power factor divided by relaxation time of MSb$_2$ (M = Ta, Nb) is calculated in the temperature range of 300--1000 K as shown in Fig. 8. The red vertical dashed line at zero eV indicates the Fermi level of the compound. The black vertical solid line at $\sim$64 and $\sim$82 meV of TaSb$_2$ and NbSb$_2$, respectively indicates the $\mu$ value, at which the transport properties are calculated to explain the experimental results. The maximum possible power factors are also calculated for \textit{p}-type and \textit{n}-type of these compounds. The maximum power factors for \textit{p}-type conduction are calculated as $\sim$1.14 and $\sim$1.74 mW m$^{-1}$ K$^{-2}$ at $\sim-$375 and $\sim-$320 meV, respectively, while these values are found to be $\sim$1.16 and $\sim$1.80 mW m$^{-1}$ K$^{-2}$ at $\sim$215 and $\sim$215 meV for \textit{n}-type of TaSb$_2$ and NbSb$_2$, respectively at 300 K. The \textit{p}-type and \textit{n}-type are confirmed from the sign of \textit{S} at the corresponding $\mu$ values of Fig. 4. The $\tau$ values are taken from the previous calculations (Inset of Fig. 6). The carrier concentrations corresponding to the maximum power factors for \textit{p}-type of TaSb$_2$ and NbSb$_2$ are calculated as $\sim$2.56 $\times$ 10$^{21}$ and $\sim$2.60 $\times$ 10$^{21}$ cm$^{-3}$, respectively, whereas for the  \textit{n}-type conduction of TaSb$_2$ and NbSb$_2$ these values are found to be $\sim$1.42 $\times$ 10$^{21}$ and $\sim$1.58 $\times$ 10$^{21}$ cm$^{-3}$, respectively. The predicted power factors at 300 K in this study are comparable with the power factors of $\sim$1--2 mW m$^{-1}$ K$^{-2}$ for Bi$_2$Te$_3$ parent compound \cite{li2006,zhao2009,fu2012}, though the power factors of BiTe-based doped compounds are reported as $\sim$4--7 mW m$^{-1}$ K$^{-2}$ in many literatures \cite{yamashita2003,yamashita2004,poudel2008,park2016} at/around the room temperature. However, a rigorous effort is required to synthesize the suitable \textit{p} and \textit{n}-type doping of MSb$_2$ (M = Ta, Nb) to validate the computational prediction.

\section{CONCLUSIONS}            
The thermoelectric properties of MSb$_2$ (M = Ta, Nb) are studied in the present study. These compounds were prepared by a combined solid-state reaction and a spark plasma sintering process. The monoclinic phase with space group \textit{C}2/\textit{m} is confirmed for both compounds through Rietveld refinement. The negative sign of the Seebeck coefficient indicates the \textit{n}-type behaviour of these compounds. The DFT-based electronic structure calculations were carried out in order to understand the experimentally observed thermoelectric properties. The semimetallic behaviour of these compounds was confirmed from the band-structure and density of states calculations. The multi-band electron and hole pockets are found reasonably good in explaining the experimental results. Further study of computational calculations gives the maximum possible power factors of $\sim$1.14 and $\sim$1.74 mW m$^{-1}$ K$^{-2}$ for \textit{p}-type conduction of TaSb$_2$ and NbSb$_2$, respectively at 300 K, while these values are found to be $\sim$1.16 and $\sim$1.80 mW m$^{-1}$ K$^{-2}$ at 300 K for \textit{n}-type conduction, respectively. In conclusion, the combined DFT and Boltzmann transport theory are found to be reasonably good at addressing the experimental transport properties, and moderate power factors can be obtained if these compounds are synthesized with the proper doping concentration.

\section{ACKNOWLEDGEMENTS} 
This work was supported by JST Mirai Program Grant Number JPMJMI19A1.

\end{document}